\documentclass[aps,twocolumn,preprintnumbers,prd,superscriptaddress,nofootinbib,10pt]{revtex4-2}

\usepackage{amssymb,amsmath,amsthm,amsfonts}
\usepackage[usenames]{color}
\definecolor{darkred}{rgb}{0.5,0,0}

\usepackage{bm}
\usepackage{dcolumn}
\usepackage[utf8]{inputenc}
\usepackage{latexsym}
\usepackage{rotating}
\usepackage{longtable}

\usepackage{enumerate}
\usepackage{tensor,multirow}
\usepackage{url}
\usepackage[linktocpage]{hyperref}
\usepackage{float}
\usepackage{graphicx}

\begin{document}

\title{Solving the Teukolsky equation with physics-informed neural networks}

\author{Raimon Luna}
	\affiliation{Departamento de Astronom\'{i}a y Astrof\'{i}sica, Universitat de Val\`{e}ncia,
Dr. Moliner 50, 46100, Burjassot (Val\`{e}ncia), Spain}

\author{Juan Calder\'on~Bustillo}
	\affiliation{Instituto Galego de F\'{i}sica de Altas Enerx\'{i}as, Universidade de
Santiago de Compostela, 15782 Santiago de Compostela, Galicia, Spain}
	\affiliation{Department of Physics, The Chinese University of Hong Kong, Shatin, N.T., Hong Kong}

\author{Juan Jos\'{e} Seoane Mart\'{i}nez}
	\affiliation{Departamento de Astronom\'{i}a y Astrof\'{i}sica, Universitat de Val\`{e}ncia,
Dr. Moliner 50, 46100, Burjassot (Val\`{e}ncia), Spain}

\author{Alejandro Torres-Forn\'e}
	\affiliation{Departamento de Astronom\'{i}a y Astrof\'{i}sica, Universitat de Val\`{e}ncia,
Dr. Moliner 50, 46100, Burjassot (Val\`{e}ncia), Spain}
  \affiliation{Observatori Astron\`{o}mic, Universitat de Val\`{e}ncia,
C/ Catedr\'{a}tico Jos\'{e} Beltr\'{a}n 2, 46980, Paterna (Val\`{e}ncia), Spain}

\author{Jos\'e A. Font}
	\affiliation{Departamento de Astronom\'{i}a y Astrof\'{i}sica, Universitat de Val\`{e}ncia,
Dr. Moliner 50, 46100, Burjassot (Val\`{e}ncia), Spain}
	\affiliation{Observatori Astron\`{o}mic, Universitat de Val\`{e}ncia,
C/ Catedr\'{a}tico Jos\'{e} Beltr\'{a}n 2, 46980, Paterna (Val\`{e}ncia), Spain}

\begin{abstract}
We use physics-informed neural networks (PINNs) to compute the first quasi-normal modes of the Kerr geometry via the Teukolsky equation. This technique allows us to extract the complex frequencies and separation constants of the equation without the need for sophisticated numerical techniques, and with an almost immediate implementation under the \texttt{PyTorch} framework. We are able to compute the oscillation frequencies and damping times for arbitrary black hole spins and masses, with accuracy typically below the percentual level as compared to the accepted values in the literature. We find that PINN-computed quasi-normal modes are indistinguishable from those obtained through existing methods at signal-to-noise ratios (SNRs) larger than 100, making the former reliable for gravitational-wave data analysis in the mid term, before the arrival of third-generation detectors like LISA or the Einstein Telescope, where SNRs of ${\cal O}(1000)$ might be achieved.
\end{abstract}

\maketitle

\section{Introduction}
\label{Introduction}

The observation of gravitational waves from binary black hole (BBH) mergers by the LIGO-Virgo-KAGRA (LVK) detector network~\cite{LIGOScientific:2016aoc, LIGOScientific:2018mvr, LIGOScientific:2020ibl,GWTC-3} has opened a window to study gravity in its strong field regime. In particular, some of the current detections have made possible to access the ringdown signal of the highly distorted black holes produced after the merger, opening the prospects to turn black-hole spectroscopy into an observational field (see e.g.~\cite{Berti:2005ys, 2018PhRvD..98j4020C, 2019PhRvD..99l3029C, 2021PhRvD.103l4043C, 2022PhRvD.105f2009C, Isi:2019aib,Bustillo2021,Capano:2105.05238}). This discipline aims to identify and characterize the quasi-normal modes (QNMs) of a rotating (Kerr) black hole, as those are the fingerprint of relaxing black holes after a perturbation of their spacetime geometry. In black-hole perturbation theory any type of perturbation (scalar, electromagnetic and gravitational) can be mapped into wave equations with a short range potential. QNMs are solutions of those equations that correspond to damped oscillating modes, i.e.~modes which are outgoing at infinity and ingoing at the black hole horizon. The properties of these modes are well understood (see~\cite{Chandra,KS99,Berti:2009kk} for details). 

A generic excitation of an isolated black hole leads to a prompt emission of gravitational waves, followed by a distinctive exponentially decaying QNM ringing. Such a perturbation can naturally arise in a variety of astrophysical situations. Examples include the late stages of the gravitational collapse of massive stars or the capture of infalling matter. With the advent of gravitational-wave astronomy and the detection of signals from BBH mergers, the QNM ringdown of the final black hole has become the prime example to test black hole perturbation theory. The ringdown can be searched for in the post-merger section of the gravitational-wave signal, which encodes information on the properties of the remnant object. In particular, the analysis of QNMs can be exploited to test the nature of such object and, specifically, to understand if this is consistent with an astrophysical black hole as predicted by general relativity, ruling out alternative exotic compact objects \cite{Cardoso:2019rvt}.

One of the most appealing prospects of black-hole spectroscopy is, thus, the testing of general relativity. By comparing the observed QNM spectra of post-merger black holes with the known spectra of the Kerr (or Kerr-Newman) solutions of Einstein(-Maxwell) theories, tests of the no-hair theorem \cite{Press1971,Vishveshwara1970,Teukolsky1973,Chandrasekhar:1975zza} may become possible. This conjecture states that for a Kerr-Newman black hole the mode oscillation frequencies and damping rates are determined uniquely by the black hole mass, angular momentum and possibly charge (which is expected to be zero for astrophysical black holes). Consistency tests of general relativity need, at least, the precise identification of three QNM parameters, two to measure the mass and the spin of the black hole and the third one for cross-check \cite{Dreyer2004,Berti2015}. Such an identification is expected to be only possible if the signal-to-noise ratio (SNR) in the ringdown stage is ${\cal O}(100)$~\cite{Berti:2005ys,Baibhav:2019,Berti:2007}. Therefore, the actual realization of strong-confidence tests of GR through black-hole spectroscopy seems out of reach for current second-generation detectors\footnote{As an example, the ringdown SNR for the first signal detected, GW150914, was around 7 when the analysis is started 3ms after the signal peak. Moreover, this reaches $\simeq 15$ if the analysis is started at the signal peak introducing overtones \cite{Isi:2019aib,Giesler2019,Bustillo2021}.} (but see~\cite{2018PhRvD..98j4020C} for a more promising estimate when various signal posteriors are {\it combined}) while it will be possible with the Einstein Telescope~\cite{ETScience} and the space detector LISA~\cite{LISA}. At present, searches for violations of the no-hair conjecture are  conducted by introducing potential deviations in the QNM parameters as additional degrees of freedom to estimate through Bayesian inference. The most recent study carried out by the LVK collaboration using data form the third gravitational-wave catalog GWTC-3~\cite{GWTC-3} constrains the frequency deviation parameters for the fundamental and the first overtone of the $\ell=m=2$ QNM by about 1\% at 90\% credibility~\cite{TGR:2021} (see~\cite{Isi:2019aib,Giesler2019} for the first application of this technique). Black hole spectroscopy is expected to become increasingly important as gravitational-wave detectors become more sensitive to subdominant quasi-normal modes. Moreover, the possible deviations from the predictions of general relativity can be parametrized and used to test for exotic compact objects \cite{Cardoso:2019rvt} and alternative theories of gravity \cite{Horndeski1974SecondorderSF, Clifton:2011jh}. 

The QNM frequencies and their decay rates are computed from the linear analysis around the background geometry, and can be extracted from a large variety of spacetime geometries (see \cite{Berti:2009kk, Cardoso:2001bb, Cardoso:2003sw, Cardoso:2008bp, Konoplya:2011qq, Yang:2012he, Pani:2013ija, Dias:2015wqa} for an incomplete list). In the case of the Schwarzschild geometry, the perturbations were first studied by \cite{Regge:1957td, Zerilli:1970se, Chandrasekhar:1975zza} and then extended to Kerr black holes by Teukolsky \cite{Teukolsky1973}, leading to the renowned master equation bearing his name. After separation of variables on the Teukolsky equation, two master equations depending respectively on the radial and zenital coordinates are obtained, which have to be solved in order to extract the QNMs.
There is a surprisingly large diversity of methods to solve such master equations. Those have been devised during the past decades in parallel to the development of computational power. The first method, introduced by Leaver \cite{10.2307/2397876} and based on the Fr\"obenius method, expands the master variables as an infinite series, imposing the convergence of such series in the whole domain through the so-called ``quotient method''. This is equivalent to finding roots of a function which is expressed as a continued fraction, which is the reason why this approach is commonly known as the ``continued fraction'' method. Among the many other methods, it is worth mentioning the direct time evolution of the equations~\cite{Krivan:1997,Pazos:2005,Harms:2013}, the computation of bound states in an effective potential \cite{Ferrari:1984zz}, the WKB approximation \cite{Iyer:1986np}, asymptotic iteration methods \cite{Cho:2011sf}, pseudospectral basis decomposition \cite{PhysRevD.90.124021, Jansen:2017oag}, or even more exotic approaches such as the expansion in the spacetime dimension \cite{Emparan:2014aba}. 

In this paper we solve the Teukolsky equation using a machine-learning approach, namely physics-informed neural networks (PINNs) \cite{2019JCoPh.378..686R} and assess the quality of our results by comparing them with those obtained by the continued fraction method. The main advantage of PINNs as a method for finding QNMs is that it can be easily adapted to different systems with very little changes in the method implementation, bringing us closer to a general-purpose quasinormal mode solver. The networks can in principle solve an arbitrary number of equations with an arbitrary number of eigenvalues and separation constants just by adding them to the loss function of the system with appropriate weighting factors. Moreover, PINNs do not require a particular computational grid or numerical differentiation techniques and the method can be easily implemented with only a few lines of code in a machine-learning environment (as e.g.~the \texttt{PyTorch} \cite{NEURIPS2019_9015} framework employed here). The equations that are to be solved have no particular requirements for the PINN to process them, as long as their solutions are finite and regular, which makes PINNs also adequate for uses on numerical background solutions.

The main idea behind PINNs is to use neural networks as universal approximants of mathematical functions \cite{HORNIK1989359} in order to solve differential equations. In the simplest case, one can approximate a function $f(x)$ with a fully-connected feed-forward neural network with a single neuron in the input layer, which receives values of $x$. Then, a single neuron (or two, if $f$ is complex) in the output layer will generate the value of $f(x)$. An important advantage with respect to other numerical techniques is the ability to evaluate the derivatives of $f(x)$ by standard automatic differentiation on every point, without the need for a numerical differentiation scheme on a fixed grid. 

Automatic differentiation in general, and backpropagation in particular, are the commonly used techniques to take derivatives of machine learning models with respect to their parameters (weights and biases), which are then used by an optimization algorithm to train the model. In our case, we use the same infrastructure of automatic differentiation to compute the derivatives of the model outputs with respect to the input coordinates of the differential equation (in the present case, the radial and angular coordinates). Given a differential equation of the form $F[x; f(x), f'(x), f''(x), ...] = 0$, then it is straightforward to define a loss function $\mathcal{L}$ as the norm of the values of $F$ on some set of points in the domain. Minimizing $\mathcal{L}$ is then equivalent to solving the differential equation for $f(x)$. 

When the differential equation has one or multiple parameters to be computed, these can be fitted in the same manner as the weights and biases of the neural network, making the practical implementation remarkably straightforward. General boundary conditions and solution normalizations can be included in the loss function (weak enforcement) or imposed on the neural network output (hard enforcement), see for instance \cite{Elastodynamics, QuantumEigenvalues, Hamiltonian}. PINNs have been used in the past for the computation of black-hole QNMs \cite{Ovgun:2019yor, Ncube:2021jfu, Cornell:2022enn} in the case of the Schwarzschild metric. In this paper we extend this approach to the more astrophysically relevant case of the Kerr metric. 

The rest of the paper is structured as follows. In Section \ref{sec:Teukolsky} we lay out our analytic treatment of the Teukolsky equation in order to make it suitable for its implementation in a PINN, which is described in Section \ref{sec:Network}. We present our results in Section \ref{sec:Results} and close with our conclusions in Section \ref{sec:Discussion}.

\section{The Teukolsky Equation}
\label{sec:Teukolsky}

In the Teukolsky equation \cite{Teukolsky1973} and its solution \cite{10.2307/2397876}, linear perturbations around the Kerr metric are encoded in a master variable $\psi$. These are classified by their spin-weight under the Lorentz group as scalar ($s = 0$), vector ($s = -1$) and tensor ($s = -2$) perturbations. In this paper we will always take $s = -2$. After performing a standard separation of variables, we obtain
\begin{equation}
\psi = \frac{1}{2\pi} \int e^{-i \omega t} \sum_{l = |s|}^\infty \sum_{m = -l}^l e^{im\phi} S_{lm}(\theta)R_{lm}(r) d\omega,
\end{equation}
where $\theta$ and $\phi$ are the zenithal and azimuthal angular coordinates, and $\omega$ is the (complex) frequency of the mode. We are left with two ordinary differential equations (ODEs) for the radial and angular parts of the solution. When solving for the modes, we will take $G = c = 2M = 1$, with $c$ denoting the speed of light in vacuum, $G$ Newton's constant, and $M$ the black hole mass. We reinstate these constants when generating waveforms for particular values of the mass $M$ and spin $a$. The radial part, for a solution with intrinsic angular momentum $J = 2Ma$, has the form
\begin{equation}
\Delta(r) R''(r) + (s+1) (2r -1)R'(r) +V(r)R(r)=0.
\label{eq:equationR}
\end{equation}
Above, we have defined
\begin{equation}
\begin{split}
\Delta(r) &= r^2 - r + a^2, \\
V(r) &= \alpha(r) \omega^2 + \beta(r) \omega + \gamma(r), \\
\alpha(r) &= \frac{(r^2+a^2)^2}{\Delta(r)} - a^2, \\
\beta(r) &= \frac{is (a^2 - r^2) - 2amr }{\Delta(r)} + 2isr, \\
\gamma(r) &= \frac{a^2 m^2 + i s a m (2r - 1)}{\Delta(r)} - A,
\end{split}
\end{equation}
where $A$ is a separation constant that will depend on the angular momentum numbers $l, m$. Furthermore, in the Schwarzschild limit  we know that \cite{10.2307/2397876}
\begin{equation}
\lim_{a \to 0}A = l(l+1) - s(s+1) .
\end{equation}
Correspondingly, the angular equation reads
\begin{equation}
\begin{split}
&(1-u^2)S''(u) - 2u S'(u) + \\ 
&\left[a^2\omega^2 u^2 - 2 a\omega s u + s + A -\frac{(m+su)^2}{1-u^2}\right]S(u) = 0,
\end{split}
\label{eq:equationS}
\end{equation}
where $u = \cos \theta$. In addition to these ODEs, any perturbation must obey ingoing boundary conditions at the horizon and outgoing at spatial infinity, as well as regularity at the spherical poles $\theta = 0, \pi$. A set of ans\"atze obeying such conditions, derived by Leaver \cite{10.2307/2397876}, are
\begin{equation}
\begin{split}
R(r) &= e^{i \omega r} (r - r_-)^{p_-} (r - r_+)^{p_+} f(x), \\
p_- &= -1-s+i\omega+i\sigma_+, \\
p_+ &= -s-i\sigma_+, \\
\sigma_+ &= \frac{\omega r_+ - a m}{\sqrt{1 - 4 a^2}}, 
\end{split}
\label{eq:definition_f}
\end{equation}
with $x = 1/r$ being used to compactify the radial domain, and
\begin{equation}
S(u) = e^{a \omega u} (1 + u)^{|m-s|/2} (1 - u)^{|m+s|/2} g(u),
\label{eq:definition_g}
\end{equation}
where $r_\pm = \left(1 \pm \sqrt{1 - 4 a^2}\right) / 2$ are respectively the inner $(r_-)$ and outer $(r_+)$ horizon radii of the Kerr metric. The functions $f(x)$ and $g(u)$ have to be regular in the ranges $[0,1]$ and $[-1,1]$ respectively. By substitution of (\ref{eq:definition_f}) and (\ref{eq:definition_g}) into (\ref{eq:equationR}) and (\ref{eq:equationS}), we obtain a set of equations for the functions $f(x)$ and $g(u)$ of the form 

\begin{equation}
\begin{split}
\mathcal{L}_F [f(x)] = F_2 f''(x) + F_1 f'(x) + F_0 f(x) =0, \\ 
\mathcal{L}_G [g(u)] = G_2 g''(u) + G_1 g'(u) + G_0 g(u) =0, \\ 
\end{split}
\label{eq:fg}
\end{equation}

\noindent
where the explicit form of the $F_i(x)$ and $G_i(u)$ coefficients is given in Appendix \ref{app:Equations}. With this, the problem reduces to finding the set of $\{\omega_n, A_n\}$ that admit nonzero solutions for (\ref{eq:fg}). Following Leaver, we choose the normalization conditions 
\begin{equation}
f(1) = 1, \quad g(-1) = 1.  
\label{eq:norm}
\end{equation}

\section{The Neural Network}
\label{sec:Network}

In order for the PINN to solve the equations, we must define two sets of values for the variables $x$ and $u$ where the networks will be evaluated. For the sake of simplicity, we pick uniformly spaced grids, as
\begin{equation}
\begin{split}
x_i = \frac{i-1}{N_x-1}, \quad i &= 1, \dots, N_x,\\
u_j = \frac{2(j-1)}{N_u-1} - 1, \quad j &= 1, \dots, N_u,
\end{split}
\end{equation}
with $N_x = N_u = 100$, although it is important to keep in mind that a uniform grid is not a requirement. We can, therefore, naturally define a loss function as the weighted sum of the averaged moduli of the right-hand-side of equations (\ref{eq:fg}), as
\begin{equation}
\begin{split}
& \mathcal{L}[\{x_i, u_j\}] = \\&= \frac{W}{N_x}\sum_{i = 0}^{N_x} \left|\mathcal{L}_F [f(x_i)]\right| + \frac{1}{N_u}\sum_{j = 0}^{N_u} \left| \mathcal{L}_G [g(u_j)]\right|,
\end{split}
\label{eq:loss}
\end{equation}
where the constant $W$ can be used to give different relative weights to the radial and angular equations. The specific value of $W$ is not very important as long as it is larger than 1, which gives a higher importance to the (more complicated) radial part. For the case at hand we chose $W = 10$. Under this setup, we will define two fully-connected neural networks, $\mathcal{N}_f$ and $\mathcal{N}_g$, to act as approximants to the solutions $f(x)$ and $g(u)$. 
\begin{figure}[thpb]
\begin{center}
\includegraphics[width=0.45\textwidth]{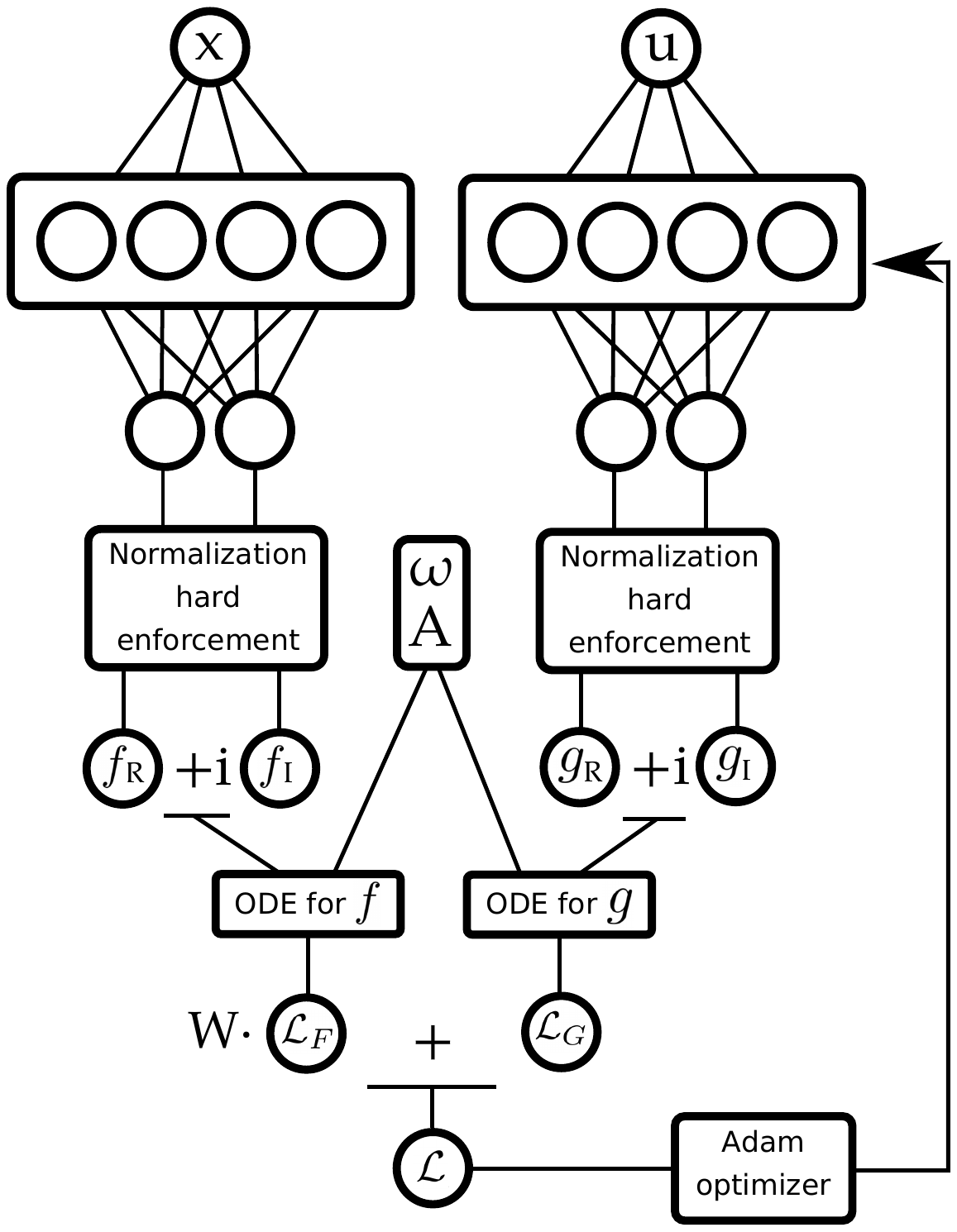}
\caption{Schematic flowchart of the structure of the neural networks. The functions $f(x)$ and $g(u)$ are evaluated independently and then introduced into the ODEs (which share the eigenvalues $\omega$ and $A$) to produce the two contributions to the loss. A weighted sum of both contributions is defined as the total loss function of the system. \label{fig:net}}
\end{center}
\end{figure}
A schematic representation of the structure of the algorithm is shown in Figure \ref{fig:net}. The input layers of both networks will contain a single neuron, which receives the value of the dummy variable ($x$ and $u$ respectively). The output layers consist of two neurons that output the real and imaginary parts of the approximated functions. The computations across both networks are performed on real numbers exclusively. The trainable parameters of the setup will then be the weights and biases of the networks and, very importantly, the parameters $\omega$ and $A$. 
Furthermore, we adopt a ``hard enforcement'' scheme to impose the normalization conditions (\ref{eq:norm}), by defining
\begin{equation}
\begin{split}
f(x) &= \left(e^{x-1} - 1\right)\mathcal{N}_f(x) + 1,\\
g(u) &= \left(e^{u+1} - 1\right)\mathcal{N}_g(u) + 1.
\end{split}
\end{equation}
This automatically ensures that (\ref{eq:norm}) are satisfied without the need for any specially devoted term in the loss (\ref{eq:loss}), which significantly accelerates the convergence of the method.

The system is implemented under the \texttt{PyTorch} framework, which provides automatic differentiation capabilities, making it very easy to implement the neural networks as well as the optimizing tools. Optimization is performed by the Adam algorithm \cite{Adam}, with an initial learning rate of 0.005 which exponentially decays with a factor 0.999 per epoch. Each value of the spin parameter $a$ is trained for a total of 2000 epochs. The gradients of the loss with respect to the parameters of both networks are computed by automatic differentiation, so both networks are trained together as a single model.

In order to compute the QNMs for a whole range of spins in $a \in \left[0,1/2\right)$ of the Kerr geometry, which can be then interpolated, we can proceed in a sequential manner. We start by training the Schwarzschild case ($a$ = 0) with the weights and biases of the network randomly initialized. Initial random weights are normally distributed with zero mean and standard deviations of 0.05 ($f$ network) and 0.01 ($g$ network), while all biases are set to zero. The initial value of $\omega$ can be set to a value close to the expected mode (in our case we choose $\omega_0 = 0.7 - 0.1i$), and $A$ is used to fix $l$, by initializing it to its analytic value at zero spin: $A_0 = l(l+1) - s(s+1)$. 

After the solution for $a=0$ has been trained, we start progressively increasing the value of $a$ in small steps, and retrain both networks to extract the QNM at the new value of $a$, until we reach the maximum value of $a = 0.4999$. At each training step, we take as initial weights and biases the ones resulting from the training at the previous step. In this way, we start the convergence process for each value of $a$ with a configuration that is closer to the optimum than a randomly initialized network. We also guarantee that the network will always converge to the mode with the same value of $l$, which would otherwise be hard to impose for $a>0$.

We find that, when gradually increasing the value of the spin parameter $a$, it is convenient to cluster the training at higher spins. In order to achieve this, we choose increasing values of $a$ as
\begin{equation}
a_k = \frac12 \left(1 - 10^{-\frac{3(k-1)}{N_a-1}} \right), \quad k = 1, \dots, N_a,
\end{equation}
with $N_a = 20$.

With this method, the initialization of the network at each value of $a$ is closer to the solution, which improves the accuracy. When the QNM frequencies have to be evaluated for a large number of different spin parameters $a$, it is convenient to perform an interpolation over the PINN results on a sparser set of values of $a$.

\subsection{Hyperparameter study}

\begin{figure}[thpb]
\begin{center}
\includegraphics[width=0.45\textwidth]{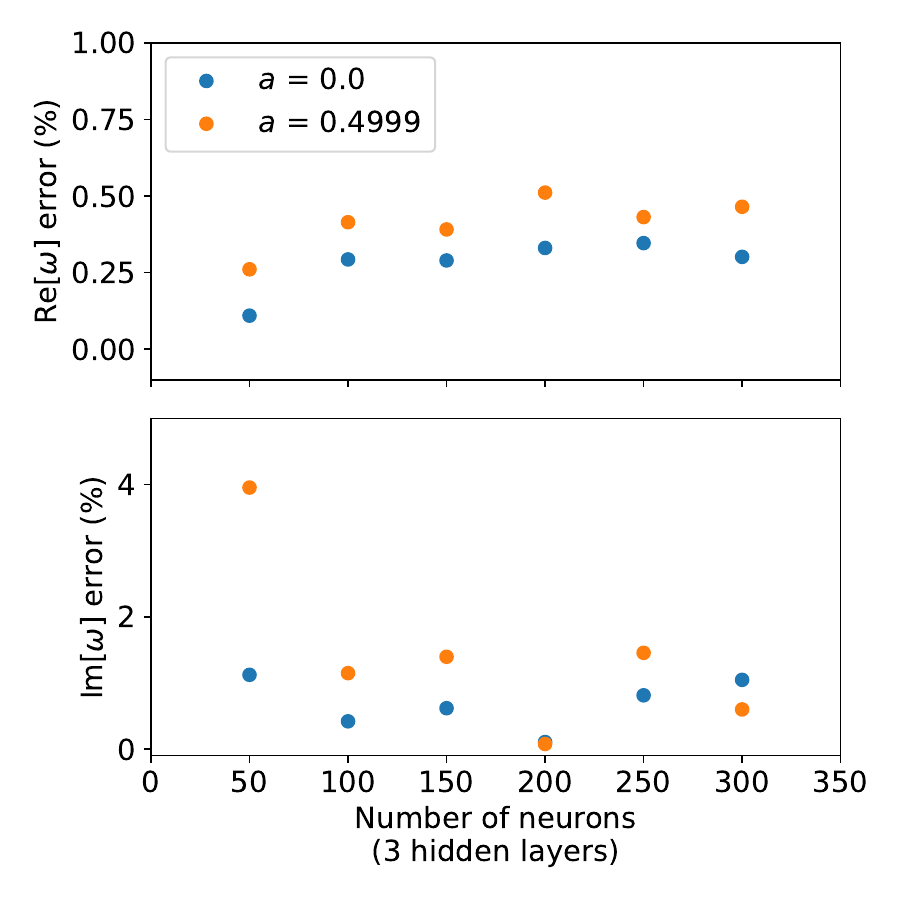}
\caption{Error of the $(l, m, n) = (2, 0, 0)$ QNM frequency for a zero spin black hole (blue circles) and a near extremal Kerr black hole (orange circles), when compared to \cite{10.2307/2397876}, as a function of the number of neurons in each of the three hidden layers. 
\label{fig:errors_neurons_per_layer}}
\end{center}
\end{figure}

\begin{figure}[thpb]
\begin{center}
\includegraphics[width=0.45\textwidth]{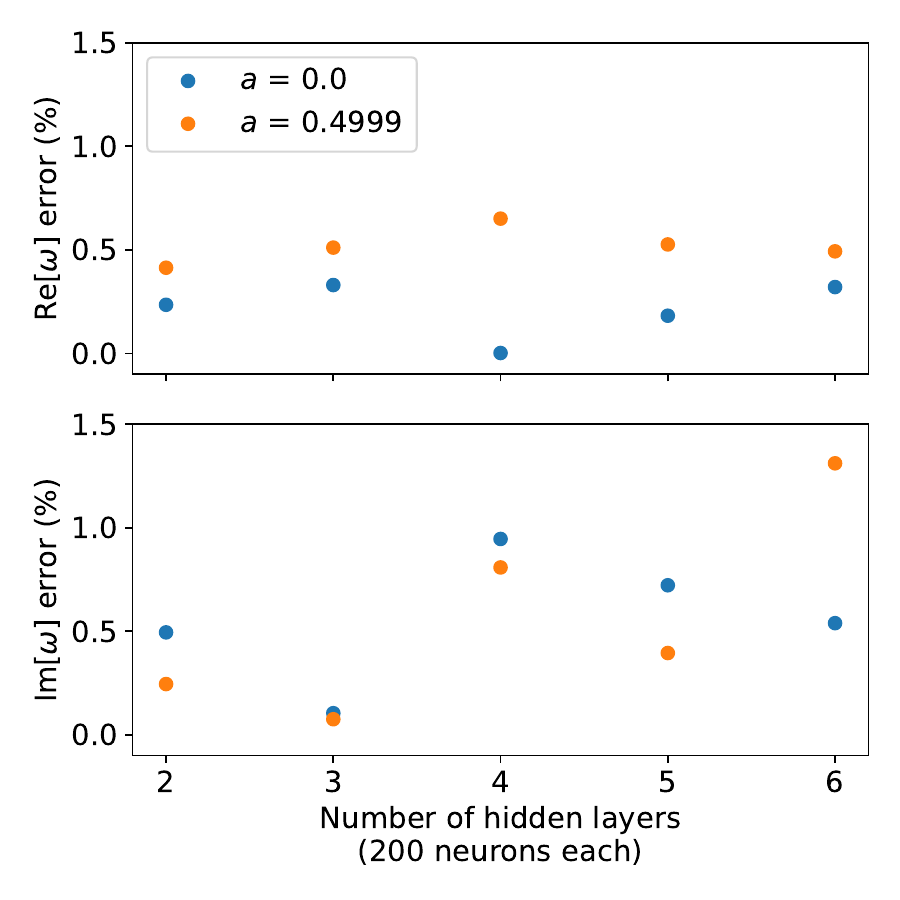}
\caption{As Fig.~\ref{fig:errors_neurons_per_layer} but showing the error of the $(l, m, n) = (2, 0, 0)$ QNM frequency as a function of the number of hidden layers, with 200 neurons each. \label{fig:errors_hidden_layers}}
\end{center}
\end{figure}

We have performed a hyperparameter study on the possible neural network architectures we could use. It revealed that moderately simple neural networks are enough for this problem. Figures \ref{fig:errors_neurons_per_layer} and \ref{fig:errors_hidden_layers} show the relative deviation of the real and imaginary parts of the mode (2,0,0) when compared to \cite{10.2307/2397876} as a function of the number of neurons per layer and the number of hidden layers, respectively. As an example we plot the results for the spinless case ($a = 0$) and the near extremal case ($a = 0.4999$). A measure of the accuracy on the actual differential equations for the radial and angular parts is the value of the loss function. For our computations, the final loss is typically of the order of $10^{-1}$.

Even though the accuracy of the results does not vary dramatically once a sufficient network complexity is achieved, for all calculations in this paper we will use 3 hidden layers of 200 neurons, with the hyperbolic tangent as the activation function. This configuration is enough to correctly obtain the values of the modes with errors below 1\%, while making the training of each mode reasonably fast: the full training of 2000 iterations takes about 2 minutes on a single Intel Core i7-10750H at 2.60GHz CPU.

%
\section{Results}
\label{sec:Results}
%
\subsection{QNM Computation}
\begin{figure}[thpb]
  \centering
  \includegraphics[width=0.5\textwidth]{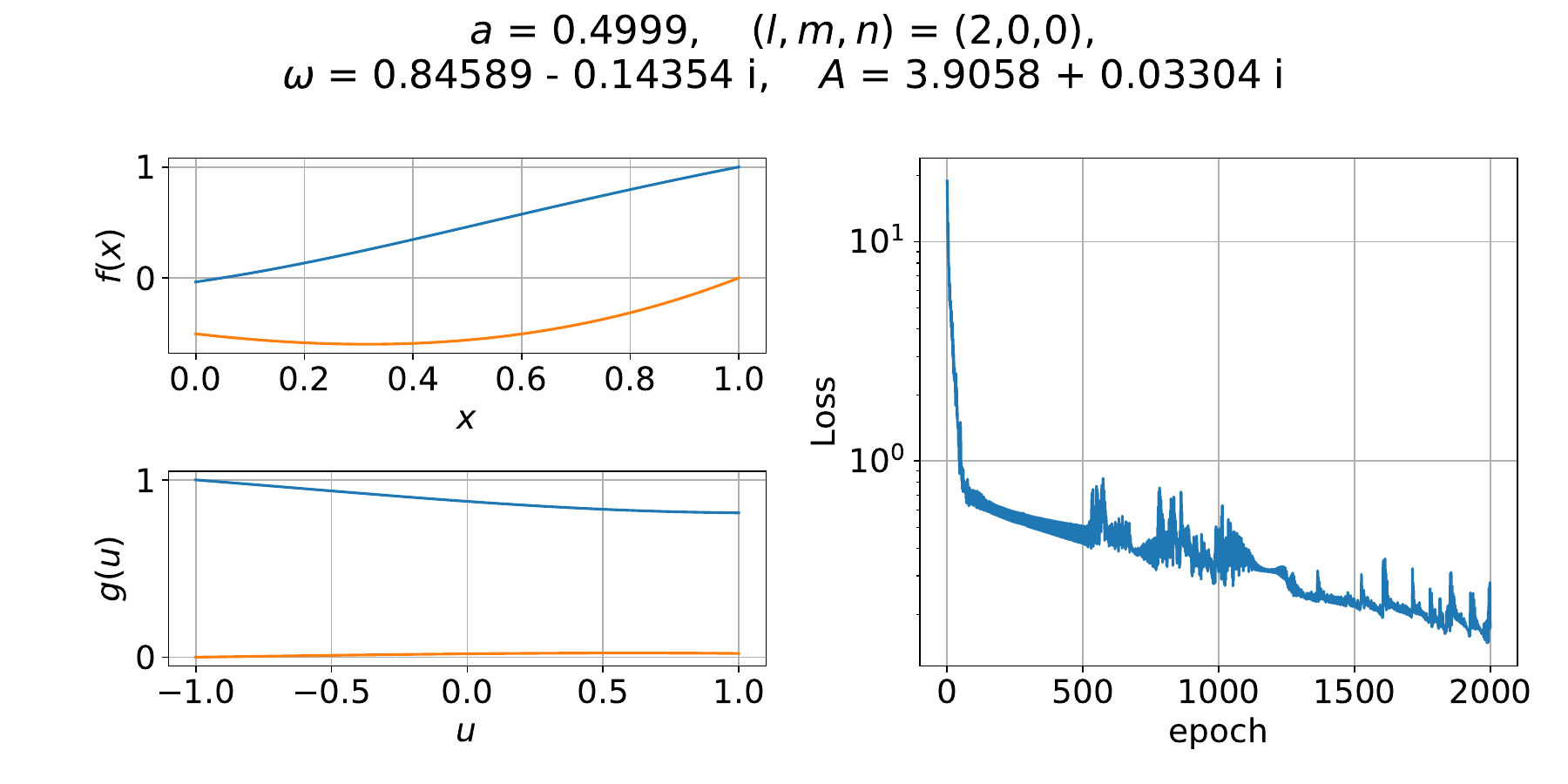}
  \caption{$f(x)$ and $g(u)$ obtained from the PINN for the dominant QNM in the case $(l, m, n) = (2, 0, 0)$, together with their learning curve, for $a$ = 0.4999. The blue and orange curves correspond to the real and imaginary parts respectively. \label{fig:learning_curves}}
\end{figure}

We use the neural networks to compute the values of the QNMs reported in \cite{10.2307/2397876}, particularly those with $(l, m, n) = (2, 2, 0)$ and $(2,0,0)$, with spin parameters ranging from $a=0$ (Schwarzschild black hole) to $a = 0.4999$. While our method is applicable to any ringdown mode, we choose to showcase these two. The first one is the fundamental emission mode from circular black-hole merger remnants. The second is co-dominant in the case of highly eccentric mergers, a type of source that has received much attention recently (e.g. \cite{Bustillo2021_headon,RomeroShaw2020,Gayathri2022,Gamba2022,Bustillo2021_proca}), and therefore may serve as a smoking gun for such scenario. In Figure \ref{fig:learning_curves} we show the final results for the near-extremality case, i.e., $a = 0.4999$ as the values of the learned radial and angular functions $f(x)$ and $g(u)$ and the learning curve for this value of the spin. The learning curve is a plot of the loss function from equation (\ref{eq:loss}) as a function of the number of iterations. As expected for a converging system, it starts by dropping very fast, and then the convergence rate becomes smaller as we approach the minimum.

The real and imaginary parts of the QNMs are explicitly reported in Tables \ref{tab:real} and \ref{tab:imag}, respectively, together with their relative deviation from the values in \cite{10.2307/2397876}. In this particular case, we randomly initialize the networks for each value of $a$. It can be observed that all of the modes in the tables are less than 1\% away.

\begin{table}[thpb]
  \centering
  \begin{tabular}{|c|c|c|c|}
    \hline
    \multicolumn{4}{|c|}{Real part of $\omega$} \\ \hline \hline
    $a$ & PINN & Leaver & \% error \\ \hline
    0 & 0.74981 & 0.74734 & 0.330 \\ \hline
    0.1 & 0.75137 & 0.75025 & 0.149 \\ \hline
    0.2 & 0.76151 & 0.75936 & 0.282 \\ \hline
    0.3 & 0.77758 & 0.77611 & 0.190 \\ \hline
    0.4 & 0.80571 & 0.80384 & 0.233 \\ \hline
    0.45 & 0.82380 & 0.82401 & 0.025 \\ \hline
    0.49 & 0.83974 & 0.84451 & 0.564 \\ \hline
    0.4999 & 0.84589 & 0.85023 & 0.511 \\ \hline
  \end{tabular}
  \caption{Numerical results obtained by the PINN for the oscillatory (real) part of the QNM $\omega$ for $(l,m,n) = (2, 0, 0)$, compared to Leaver's results. The error lies below the percentual level in all cases. \label{tab:real}}
\end{table}
\begin{table}[thpb]
  \centering
  \begin{tabular}{|c|c|c|c|}
    \hline
    \multicolumn{4}{|c|}{Imaginary part of $\omega$} \\ \hline \hline
    $a$ & PINN & Leaver & \% error \\ \hline
    0 & -0.17812 & -0.17793 & 0.108 \\ \hline
    0.1 & -0.17719 & -0.17740 & 0.122 \\ \hline
    0.2 & -0.17610 & -0.17565 & 0.255 \\ \hline
    0.3 & -0.17332 & -0.17199 & 0.775 \\ \hline
    0.4 & -0.16572 & -0.16431 & 0.853 \\ \hline
    0.45 & -0.15785 & -0.15697 & 0.566 \\ \hline
    0.49 & -0.14657 & -0.14707 & 0.337 \\ \hline
    0.4999 & -0.14354 & -0.14365 & 0.073 \\ \hline
  \end{tabular}
  \caption{Numerical results obtained by the PINN for the damping (imaginary) part of the QNM $\omega$ for $(l,m,n) = (2, 0, 0)$, compared to Leaver's results. The error lies below the percentual level in all cases. \label{tab:imag}}
\end{table}
Subdominant QNMs can also be computed with the PINN, although higher overtones ($n > 0$) become increasingly difficult to obtain and usually require larger training times. The simplest approach is to initialize the frequency parameter $\omega$ to a value closer to the desired mode prior to the training, which can make the training process converge to it. Slowly lowering the initial imaginary part of $\omega$ can be used as a way to scan the first overtones. More sophisticated methods use driving terms in the loss function to favor higher modes, as in \cite{QuantumEigenvalues}. As an illustration, in Table \ref{tab:subdominant} we show the  subdominant modes $(3,3,0)$ and $(2,2,1)$ for the spinless case ($a=0$), together with their errors\footnote{Due to a slower convergence, the training of the $(2,2,1)$ mode needed 5000 iterations instead of 2000, and required the initial guess for $\omega$ to be $0.7-0.5 i$, quite close to the correct value.}. These two QNMs are considered the best candidates for black hole spectroscopy \cite{Capano:2105.05238,Isi:2019aib,Cabero2020,Bhagwat2022}, as the $(3,3,0)$ is the loudest and longest-lived subdominant mode in BBH mergers and the $(2,2,1)$ mode is the strongest overtone when only the spherical mode $(2,2)$ is observable in the signal during the merger stage, as it is the case for all GW detections to date (see however \cite{Capano:2105.05238}). For the sake of simplicity, for the rest this paper we will restrict to the fundamental $n=0$ modes.

\begin{table}[thpb]
  \centering
  \begin{tabular}{|c|c|c|c|}
    \hline
    \multicolumn{4}{|c|}{Subdominant mode examples} \\ \hline \hline
    $(l,m,n)$ & PINN $\omega$ & $\delta$ Re & $\delta$ Im  \\ \hline
    $(3,3,0)$ & $1.2016 - 0.1852 i$ & 0.23 \% & 0.11 \% \\ \hline
    $(2,2,1)$ & $0.6943	- 0.5490 i$ & 0.13 \% & 0.21 \% \\ \hline
  \end{tabular}
  \caption{Subdominant modes obtained by the PINN for the Schwarzschild black hole, together with their errors with respect to the results in \cite{10.2307/2397876}. \label{tab:subdominant}}
\end{table}

\subsection{Detectability Prospects}
\begin{figure}[thpb]
\begin{center}
\includegraphics[width=0.5\textwidth]{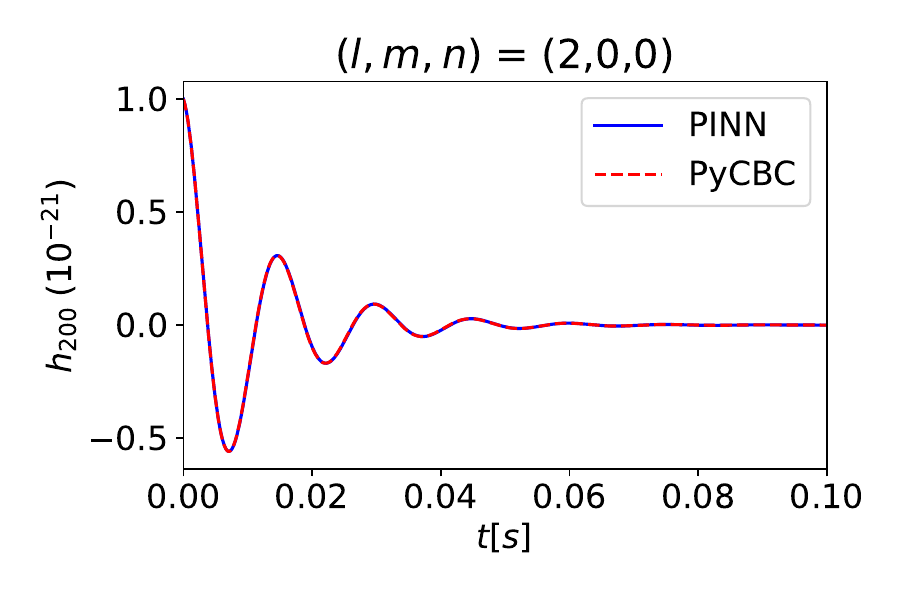}
\includegraphics[width=0.5\textwidth]{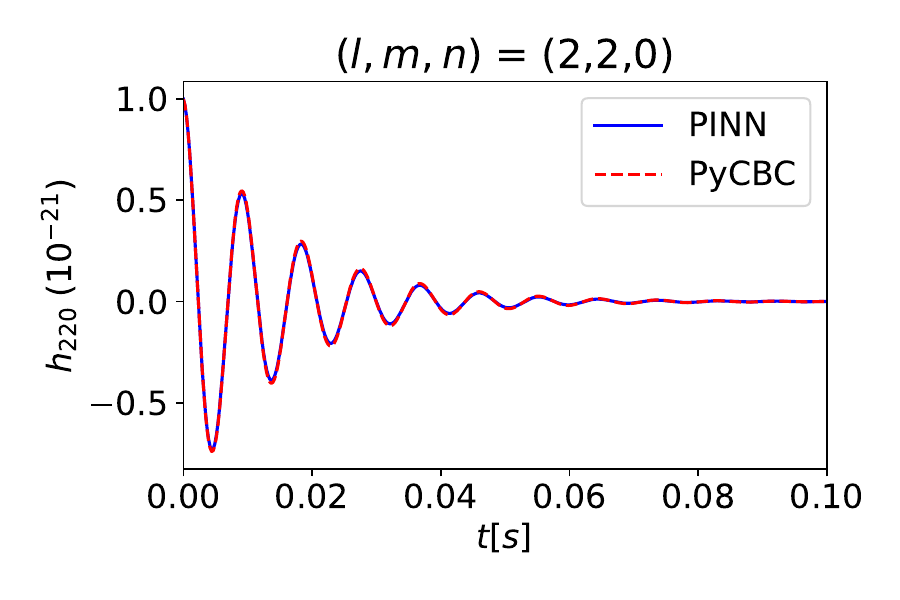}
\caption{Time-domain waveform comparison between the $h_{200}$ (top) and $h_{220}$ (bottom) strain predictions of the PINN and the reference by \texttt{pykerr}, for a black hole of mass $M=200 M_\odot$ and angular momentum $J = 0.9M$. \label{fig:Waveform}}
\end{center}
\end{figure}

In order to assess the quality of the QNMs generated from the PINN for parameter estimation (mass, spin, amplitude and phase) of a ringing astrophysical black hole detection, we can produce ringdown waveforms as damped sinusoid strain signals of the form
\begin{equation}
h = h_0 e^{-t/\tau} \cos(2 \pi f t + \phi),
\label{eq:sinusoid}
\end{equation}
 for every set of $(l,m,n)$. Above, $h_0$ and $\phi$ are an initial amplitude and phase determined by the particular perturbation, which in Figure \ref{fig:Waveform} we choose to be $10^{-21}$ and 0 respectively. The frequency $f$ and damping rate $\tau$ for a black hole of mass $M$ are computed respectively from the real and imaginary parts of $\omega$ as
\begin{equation}
\begin{split}
f &= \frac{c^3}{4 \pi G M} \text{Re}[\omega],\\
\tau &= - \frac{2 G M}{c^3} \frac{1}{\text{Im}[\omega]}\,.
\end{split}
\end{equation}
%
%
\begin{figure}[thpb]
\begin{center}
\includegraphics[width=0.5\textwidth]{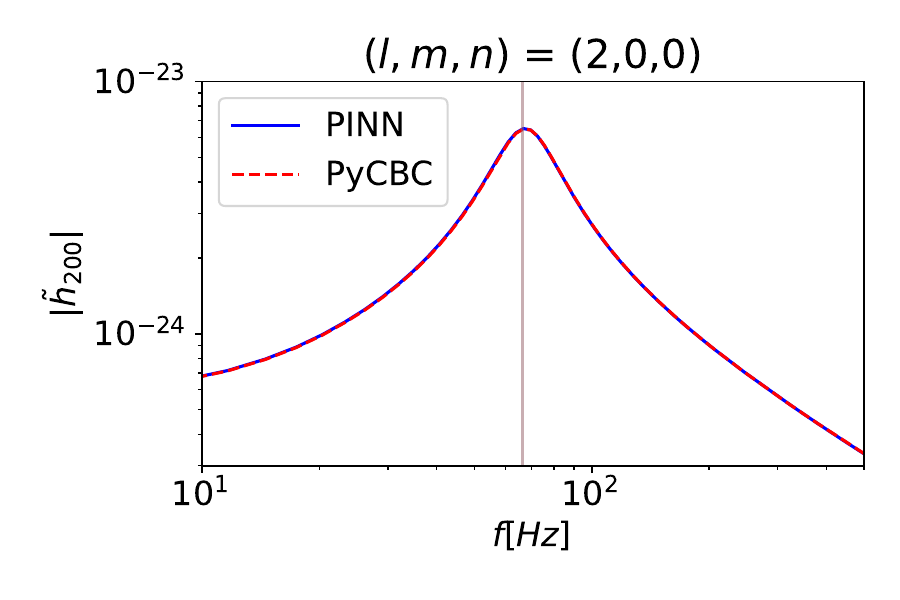}
\includegraphics[width=0.5\textwidth]{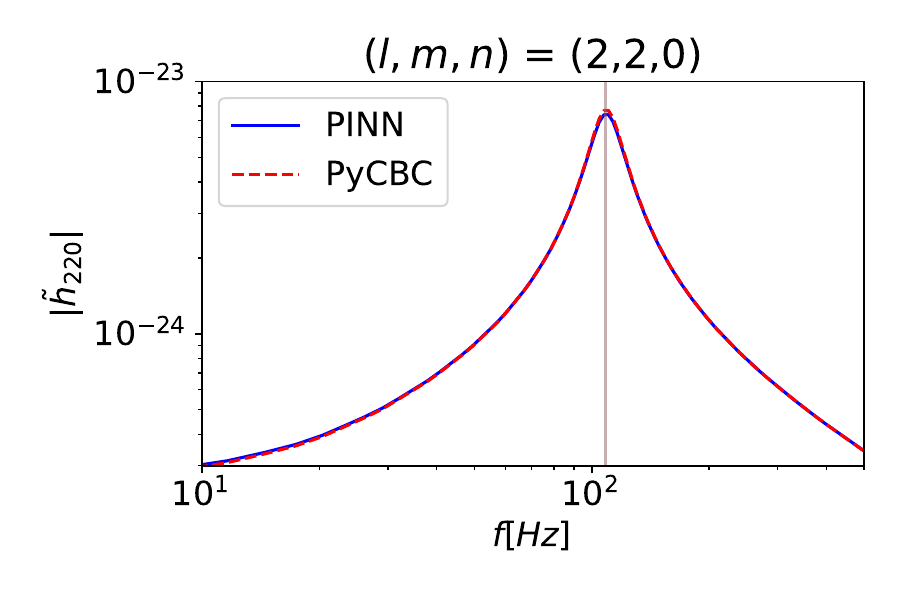}
\caption{Fourier transform of the $h_{200}$ (top) and $h_{220}$ (bottom) strain predictions, for a black hole of mass $M=200 M_\odot$ and angular momentum $J = 0.9M$. The vertical lines indicate the predictions by the PINN and by PyCBC. However, they cannot be told apart in the figure, since their frequencies only differ by 0.17 Hz and 0.07 Hz respectively. \label{fig:Fourier}}
\end{center}
\end{figure}
For each mass and spin configuration, we produce two waveforms using the $\omega$ values respectively computed by the PINN and the Python library \texttt{pykerr}, which uses the tabulated values from \cite{Berti:2005ys} available at the GRIT Ringdown website \cite{GRIT}.

The top and bottom panels of 
Fig.~\ref{fig:Waveform} respectively show the (2,0,0) and (2,2,0) waveforms obtained for the case of a black hole with parameters $M = 200M_\odot$ and $J = 0.9M$ in the time domain, while the corresponding Fourier transformed waveforms are shown in 
Fig.~\ref{fig:Fourier}. 

We quantify the discrepancy between the two waveforms by means of their match \cite{Cutler:1994ys, Apostolatos:1995pj, Finn:1992wt}, through the library \texttt{PyCBC} \cite{alex_nitz_2022_6912865}.
\begin{figure}[thpb]
\begin{center}
\includegraphics[width=0.45\textwidth]{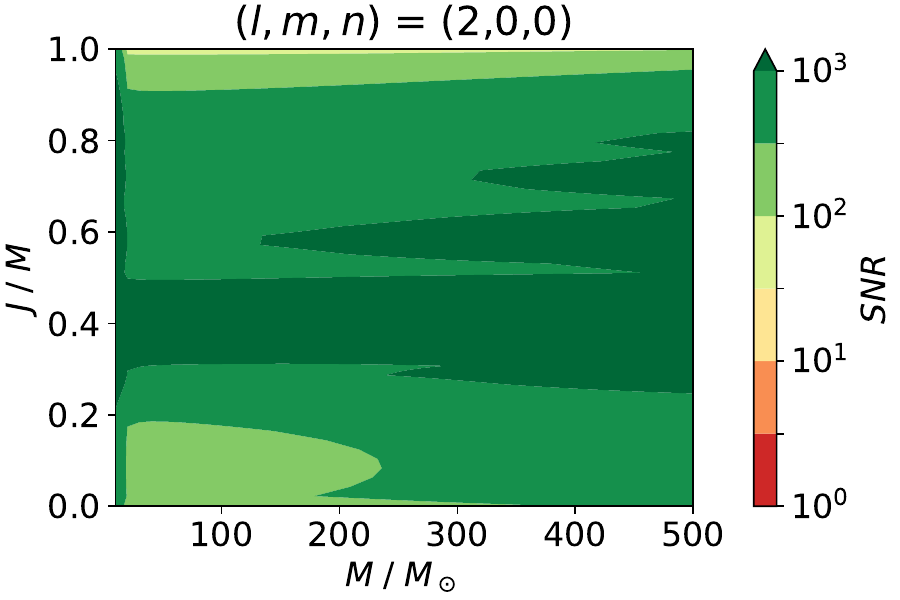}
\includegraphics[width=0.45\textwidth]{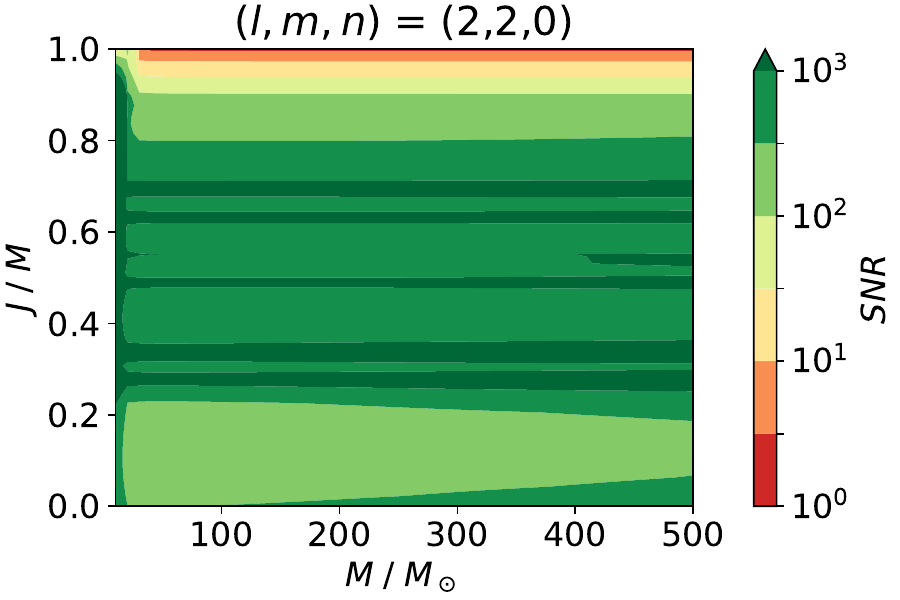}
\caption{Signal-to-noise ratio estimates to resolve the waveforms generated by using modes from the PINN and \texttt{pykerr}.\label{fig:SNR}}
\end{center}
\end{figure}
The match can be used to estimate the SNR needed to distinguish the corresponding waveforms or, in other words, the signal loudness required for the difference between the two waveforms to drive parameter biases in a real detection scenario. This is given by \cite{Lindblom2008,Chatziioannou2017,Chatziioannou:2016ezg}
\begin{equation}
F = 1 - \frac{D}{2\; \text{SNR}_{\rm min}^2},
\end{equation}
where $F$ is the match between the signals and $D$ is the number of intrinsic parameters of the model. In our case these are $D=4$: mass, spin, amplitude and phase. The match of two waveforms $a$ and $b$ is defined as their overlap maximized over time and phase shifts as
\begin{equation}
  F(a,b) = \max_{t_c,\phi}\frac{(a|b)}{\sqrt{(a|a)(b|b)}},
\end{equation}
where $(a,b)$ denotes the inner product defined as
\begin{equation}
  (a|b) = 4 \times {\rm Re}\int_{f_0}^{f_{\rm max}} \frac{\tilde{a}(f)\tilde{b}^{*}(f)}{S_n(f)}\,df\,,
\end{equation}
with $\tilde a(f)$ denoting the Fourier transform of $a(t)$, and $\tilde a^*(f)$ its complex conjugate. In addition, $S_n(f)$ is the one-sided noise power spectral density (PSD). In our case we use the \texttt{aLIGOZeroDetHighPower} model from the \texttt{LALSimulation} package \cite{lalsuite}, representative of the design sensitivity of Advanced LIGO. In the calculation of the match $F$, we numerically maximize the match over possible values of the phase difference in Eq.~(\ref{eq:sinusoid}). Therefore, we can argue that the QNM extracted by the PINNs could be used for experimental parameter estimation as long as the SNR produced by the ringdown signal in the detector is lower than our estimation. 
Fig. \ref{fig:SNR} shows the SNR$_{\rm min}$ as a function of the black hole mass and spin. We assume a detection window of 300 ms with a sampling rate of 4096 Hz. We observe that SNR$_{\rm min}$ well above 100 are obtained for spins lower than $\simeq 0.9$, typical of LIGO-Virgo observations. This indicates that our QNMs are suitable for parameter estimation in the near future, before the arrival of third-generation detectors like LISA or Einstein Telescope, when SNRs of ${\cal O}(1000)$ can be expected \cite{Bhagwat2022}. We note that the quality of the PINN modes significantly decays for higher values of the spin, as the black hole approaches extremality, especially for the case of the $(l, m, n) = (2,2,0)$ mode. 

\section{Discussion}
\label{sec:Discussion}

We have used physics-informed neural networks~\cite{2019JCoPh.378..686R} to solve the Teukolsky master equation and compute the first quasi-normal modes of the Kerr geometry for arbitrary black-hole masses and spins. Our results show an accuracy typically below the percentual level as compared to the accepted values in the literature~\cite{10.2307/2397876}. This validates the method as an acceptable approach to use for black-hole spectroscopy in the foreseeable future. We have focused here on the cases of the fundamental $(\ell,m,n)=(2,2,0)$ ringdown mode, dominant for quasi-circular BBH merger remnants and the $(2,0,0)$ mode, which can serve as a smoking gun for highly eccentric mergers \cite{Bustillo2021_proca}. In addition, we have displayed the calculation of the overtone $(2,2,1)$ and the strongest higher angular mode in quasi-circular BBH mergers $(3,3,0)$, as these are the most promising prospects to realise black-hole spectroscopy in the near future. While in this paper we have showcased the utility of PINNs in solving the astrophysically relevant case of the Teukolsky equation, this work can be regarded as a first step towards the computation of QNMs under more generic conditions (e.g.~\cite{Cheung2021}).

PINNs are very versatile, general-purpose tools that, as the particular physics case discussed in this work has shown, can solve in a remarkably straightforward way quasi-normal mode (eigenvalue) problems that would otherwise require a more involved numerical setup. Their implementation is very simple when compared to other techniques, requiring only a few lines of code in a suitable machine learning framework providing automatic differentiation capabilities. In addition, the method has also the advantage of being able to solve several equations with several eigenvalues at once, in our case the radial and angular equations with the eigenvalue $\omega$ and the separation constant $A$. As opposed to other methods, this generalization is trivially done by performing a weighted sum of each equation's contribution to the total loss function.

Moreover, the requirements on the type of equations to be used are very low, and PINNs are expected to solve very generic equations. In particular, they can be applied to any system of the form (\ref{eq:fg}) with arbitrary coefficients $F_i(x), \, G_i(u)$. Remarkably, these coefficients do not even need to have an analytical expression, which makes this approach especially interesting for the computation of QNMs for complicated numerical background solutions, such as those involving exotic compact objects or in modified gravity. Therefore, we regard this paper as a preliminary step towards the use of PINNs for the extraction of QNMs in a variety of more complex scenarios in strong-field gravity.

The major drawbacks of a solution procedure based on PINNs are the computation time, which is usually larger than in other methods, and the accuracy of the results. However, for the physics case studied in this work these two drawbacks do not seem to be much of a concern. On the one hand, the very small size of the required network architectures makes the training affordably fast, with approximately 2 minutes per mode. On the other hand, the precision that we obtain, albeit not as high as that achieved by some other numerical techniques (e.g.~spectral methods), is sufficient to be used on experimental detection data for black-hole spectroscopy and parameter estimation for current detector sensitivities, before the arrival of third-generation detectors like LISA or Einstein Telescope.

\begin{acknowledgments}

We thank Daniela Doneva and Stoytcho Yazadjiev for encouraging us to write this article. Also, we would like to thank the anonymous referee for very interesting comments and suggestions for future work.
RL acknowledges financial support provided by Next Generation EU through a University of Barcelona Margarita Salas grant from the Spanish Ministry of Universities under the {\it Plan de Recuperaci\'on, Transformaci\'on y Resiliencia} and by Generalitat Valenciana / Fons Social Europeu through APOSTD 2022 post-doctoral grant CIAPOS/2021/150.
JCB is supported by a fellowship from ``la Caixa'' Foundation (ID
100010434) and from the European Union’s Horizon 2020 research and innovation programme under the Marie Skłodowska-Curie grant agreement No 847648. The fellowship code is LCF/BQ/PI20/11760016. JCB is also supported by the research grant PID2020-118635GB-I00 from the Spain-Ministerio de Ciencia e Innovaci\'{o}n.
ATF and JAF are supported by the Spanish Agencia Estatal de Investigaci\'on (Grants PGC2018-095984-B-I00 and PID2021-125485NB-C21) funded by MCIN/AEI/10.13039/501100011033 and ERDF A way of making Europe, and the Generalitat Valenciana (Grant PROMETEO/2019/071). JAF is further supported by the European Union’s Horizon 2020 research and innovation (RISE) programme H2020-MSCA-RISE-2017 Grant No. FunFiCO-777740.

\end{acknowledgments}

\hfill
\appendix

\section{PINN equations}
\label{app:Equations}

The coefficients for equations (\ref{eq:fg}) that are actually solved by the PINN are
\begin{widetext}
\begin{equation}
\begin{split}
F_0(x) &= -a^4 x^2 \omega ^2-2 a^3 m x^2 \omega \\ & + a^2 \left(-A x^2 + x^2 \left(4 \left(r_++1\right) \omega ^2+2 i \left(r_++2\right) \omega +2 i s (\omega +i)-2\right)+ x \omega ^2- \omega ^2\right) \\ & +2 a m \left(r_+ x^2 (2 \omega +i)-x (\omega +i)-\omega \right)+A (x-1) \\ & -i r_+ (2 \omega +i) \left(x^2 (s-2 i \omega +1)-2 (s+1) x+2 i \omega \right)+(s+1) (x-2 i \omega ),\\
F_1(x) &= 2 a^4 x^4 (x-i \omega )-2 i a^3 m x^4+a^2 x^2 \left(2 r_+ x^2 (-1+2 i \omega )-(s+3) x^2+2 x (s+i \omega +2)-4 i \omega \right) \\ & +2 i a m (x-1) x^2+(x-1) \left(2 r_+ x^2 (1-2 i \omega )+(s+1) x^2-2 (s+1) x+2 i \omega \right),\\
F_2(x) &= a^4 x^6-2 a^2 (x-1) x^4+(x-1)^2 x^2,
\end{split}
\end{equation}
\begin{equation}
\begin{split}
G_0(u) &= 4 a^2 \left(u^2-1\right) \omega ^2-4 a \left(u^2-1\right) \omega  ((u-1) \left| m-s\right| +(u+1) \left| m+s\right| +2 (s+1) u) \\ & +4 \left(A \left(u^2-1\right)+m^2+2 m s u+s \left((s+1) u^2-1\right)\right)-2 \left(u^2-1\right) \left| m+s\right| \\ & -2 \left(u^2-1\right) \left| m-s\right|  (\left| m+s\right| +1)-(u-1)^2 \left| m-s\right| ^2-(u+1)^2 \left| m+s\right| ^2,\\
G_1(u) &= -8 a \left(u^2-1\right)^2 \omega -4 \left(u^2-1\right) ((u-1) \left| m-s\right| +(u+1) \left| m+s\right| +2 u),\\
G_2(u) &= -4 \left(u^2-1\right)^2.
\end{split}
\end{equation}

\end{widetext}

\bibliography{references}

\end{document}